\documentclass[12pt,reqno,a4paper]{article}
\usepackage{amsmath,amsfonts,amssymb,graphicx,epsfig,pict2e,rotating}
\usepackage[english]{babel}
\usepackage[noxcolor]{pstricks}
\usepackage{hyperref}
\numberwithin{equation}{section}

\newcommand{\be}{\begin{equation}}
\newcommand{\ee}{\end{equation}}
\newcommand{\bea}{\begin{eqnarray}}
\newcommand{\eea}{\end{eqnarray}}
\newcommand{\D}{{\mathbf D}}
\newcommand{\A}{{\mathbf A}}
\newcommand{\HH}{{\mathbf H}}
\newcommand{\B}{{\mathbf B}}

\newcommand{\uno}{{\mathbf 1}}
\newcommand{\T}{{\mathbf T}}
\newcommand{\e}{{\mathbf e}}

\begin{document}
\setcounter{page}{1}

\vspace{8mm}
\begin{center}
{\Large {\bf Lattice Integrals of Motion of the Ising Model on the Strip  }}

\vspace{10mm}
 { Alessandro Nigro\\
 Dipartimento di Fisica and INFN- Sezione di Milano\\
 Universit\`a degli Studi di Milano I\\
 Via Celoria 16, I-20133 Milano, Italy\\
 Alessandro.Nigro@mi.infn.it}\\
 [.4cm]

  \end{center}

\vspace{8mm}
\centerline{{\bf{Abstract}}}
\vskip.4cm
\noindent
We consider the 2D  critical Ising model on a strip with fixed boundary conditions. It is shown that for a suitable reparametrization of the known Boltzmann weights the transfer matrix becomes a polynomial in the variable $\csc(4u)$, being $u$ the spectral parameter. The coefficients of this polynomial are decomposed on the fixed boundaries Temperley-Lieb Algebra by introducing a lattice version of the Local Integrals of Motion. 

\renewcommand{\thefootnote}{\arabic{footnote}}
\setcounter{footnote}{0}
\section{Introduction}
The objective of this paper is to extend the results of the previous work on Critical Dense Polymers \cite{nigropol} and Ising on a cylinder \cite{nigroperiodic} to the case of Ising on a strip, in order to put the previous work on a more solid ground and to show that once more the same approach is possible.\\
The approach goes all the way back to \cite{blz}, where the integrable structure of conformal field theory (CFT) was analyzed from the point of view of the existence of infinitely many higher spin conserved quantities in mutual involution. Inspired by this work we seek and succeed to obtain the tower of commuting conserved quantities directly on the lattice.\\
The involutive charges of CFT are defined as the maximal Abelian subalgebra of the enveloping algebra $UVir$ of the Virasoro algebra, the lattice analogue of this structure is obtained by replacing the Virasoro algebra with the Temperley-Lieb algebra, as it is well known \cite{saleur}.\\
There are of course different types of Temperley-Lieb (TL) algebras, in particular it makes a lot of difference to work with open boundaries or periodic boundaries \cite{nigroperiodic}. In \cite{nigropol} we worked with an open boundary TL algebra, and obtained closed form expressions for many of the involutive charges, an ansatz closed form for the generic expressions was also provided in which some of the coefficients are still unfortunately not known. The heart of the difficulty, as pointed out long ago also in \cite{saleur} lies in the presence of boundary terms which manifest themselves through the appearence of boundary tangles. This should be the case also for the Ising model on a strip, and infact we succeed in finding the same type of contribution from the boundary as we did for Polymers  on a strip \cite{solvpol}\cite{nigropol}.We also succeed in finding a larger number of conserved charges on the lattice than we did for Polymers on the strip. \\

\section{Generalities}
It is well known that the Ising model can be characterized in terms of Boltzmann weights, which can be derived from an elementary lattice hamiltonian \cite{ret1}\cite{baxter}. Naturally such weights can be suitably rescaled by multiplying for trigonometric functions, in particular we shall employ here the following parametrization:
\begin{equation}\begin{split}W_R\left(\left.\begin{matrix}\rho&\\[-2mm]&\tau\\[-2mm]\sigma&\end{matrix}
\right| \,u \right)= \Big(\tan\Big(\frac{\pi}{4}-u\Big)\delta_{\sigma,\rho}+\delta_{\sigma,-\rho}\Big)\delta_{\sigma,-\tau}+\Big(\cot\Big(\frac{\pi}{4}-u\Big)\delta_{\sigma,\rho}+\delta_{\sigma,-\rho}\Big)\delta_{\sigma,\tau} \end{split} \end{equation}
\begin{equation}\begin{split}W_L\left(\left.\begin{matrix}&\rho\\[-2mm]\tau&\\[-2mm]&\sigma\end{matrix}
\right| \,u \right)= \Big(\cot(u)\delta_{\sigma,\rho}+\delta_{\sigma,-\rho}\Big)\delta_{\sigma,\tau}+\Big(\tan(u)\delta_{\sigma,\rho}+\delta_{\sigma,-\rho}\Big)\delta_{\sigma,-\tau}  \end{split} \end{equation}
satisfying
\be W_R\left(\left.\begin{matrix}\rho&\\[-2mm]&\tau\\[-2mm]\sigma&\end{matrix}
\right| \,\frac{\pi}{4}-u \right)=W_L\left(\left.\begin{matrix}&\rho\\[-2mm]\tau&\\[-2mm]&\sigma\end{matrix}
\right| \,u \right)   \ee
A transfer matrix can then be defined in different ways and for different boundary conditions, in particular if we choose to work on the strip we can build the transfer matrix in a boundary condition dependent way ,therefore we introduce the open boundaries transfer matrix:
\be T_{\mathbf{ \sigma},\mathbf{ \rho}}(u)=\sum_{{\bf \tau}}W_R\left(\left.\begin{matrix}1&\\[-2mm]&\tau_{1}\\[-2mm]1&\end{matrix}
\right| \,u \right) \prod_{n=2}^{L+1} W_L\left(\left.\begin{matrix}&\rho_n\\[-2mm]\tau_{n-1} &\\[-2mm]&\sigma_n\end{matrix}
\right| \,u \right)W_R\left(\left.\begin{matrix}\rho_n &\\[-2mm]&\tau_{n}\\[-2mm]\sigma_n&\end{matrix}
\right| \,u \right) W_L\left(\left.\begin{matrix}&b\\[-2mm]\tau_{L+1} &\\[-2mm]&b\end{matrix}
\right| \,u \right)   \ee
with 
\be \rho_{1}=1   \ee
\be \rho_{L+2}=b   \ee
and similar conditions for $\sigma$. The transfer matrix has thus size $2^L\times 2^L$.\\
The partition function is then defined as:
\be  Z=\textrm{Tr}\T^M   \ee
on general grounds the partition function is expected to behave asymptotically as a linear combination of CFT characters, whose modular parameter is related to the exponential of some trigonometric function of the spectral parameter. We will describe this behaviour of the partition function in a later section.\\
It is also well known that the Ising model and in generally Spin-chains, Potts models, RSOS models, Vertex models \cite{saleur} are directly related to representations of the Temperley-Lieb (TL) algebra. We will extensively investigate this connection for this special case, thus for our needs it is necessary to introduce the TL algebra, whose generators are defined by the following matrix elements: 
\be (\e_{2i})_{\mathbf{\sigma},\mathbf{\rho}}=\sqrt{2}\delta_{\sigma_i,\sigma_{i+1}}\prod_{k=1}^{L+2} \delta_{\sigma_k,\rho_k}  \ee
\be (\e_{2i-1})_{\mathbf{\sigma},\mathbf{\rho}}=\frac{1}{\sqrt{2}}\prod_{k\neq i}\delta_{\sigma_k,\rho_k}  \ee
which satisfy the relations:
\be \e_i^2=\sqrt{2}\e_i  \ee
\be  \e_i\e_{i\pm 1}\e_i=\e_i      \ee
\be \e_i\e_j=\e_j\e_i \ , \ |i-j|\geq 2  \ee
notice that there are $2L+2$ generators, due to the fact that the spin streaks are extended on the right and on the left boundary.

\section{Inversion Identity and Eigenvalues}
In this section we will solve the inversion identity for the transfer matrix thus obtaining the full spectrum of its eigenvalues. The approach here goes all the way back to \cite{baxter}, and more recently to \cite{ret1}.\\
The inversion identity takes the following form:
\be  \D(x)\D(-x)=(-1)^{L+1}2^{2L+4}U_{2L+2}(x)\uno \ee
where $x=\csc(4u)$ and $U_n(x)$ are the second kind Chebyshev polynomials.
The $U_{2n}$ can be factorized as:
\be U_{2L+2}=2^{2L+2}\prod_{k=1}^{L+1}\Big(x^2-\sin^2\Big(\frac{(2k-1)\pi}{4L+6}\Big)\Big)   \ee
\be  \D(x)\D(-x)=(-1)^{L+1}2^{4L+6}\prod_{k=1}^{L+1}\Big(x^2-\sin^2\Big(\frac{(2k-1)\pi}{4L+6}\Big)\Big) \ee
we then define the rescaled transfer matrix
\be \D(x)=2^{2L+3}x^{L+1}\T(x)\ee
hence the eigenvalues become
\be T(x)= \frac{1}{x^{L+1}}\prod_{k=1}^{L+1}\Big(x-\mu_k\sin\Big( \frac{(2k-1)\pi}{4L+6}\Big)\Big)\  \ee
where $\mu=(\mu_1,\mu_2...\mu_L,\mu_{L+1})$ is such that $\mu_i=\pm 1$
one can alternatively introduce the partition $\mathcal{P}$ which is made up by the $k$ such that $\mu_k=-1$ and write
\be  T(x)=\prod_{k\in\mathcal{P}}\frac{x+\sin(t_k)}{x-\sin(t_k)}\prod_{k=1}^{L+1}\Big(1-\frac{\sin(t_k)}{x}\Big) \ee 
where
\be t_k=\frac{(2k-1)\pi}{4L+6} \ee
The solutions of the inversion identity however, are too many, namely $2^{L+1}$ vs the size of the transfer matrix which is $2^L$. If we call $p$ the length of the partition ${\cal P}$ we have that for $b=+1$ we must consider $p$ as even, whereas for $b=-1$ we must take $p$ to be odd. This completes the discussion of the selection rules.

\section{Temperley Lieb Decomposition}
In this section we want to decompose the transfer matrix in terms of the Temperley-Lieb algebra as:
\be \D(x)=\sum_{k=0}^{L+1}2^{2L+3-k}\frac{\D_k}{k!}x^{L+1-k}  \ee
and for the rescaled transfer matrix
\be \T(x)=\sum_{k=0}^{L+1}\frac{\D_k}{k!(2x)^{k}}  \ee
and taking the logarithm and expanding as $x\to\infty$ we define what we shall call the lattice integrals of motion ${\bf A}_{2n-1}$
\be  \log\T(x)=\sum_{k=1}^{\infty}\frac{\A_k}{k!(2x)^{k}}  \ee
being
\be  \D_k=B_k(\A_1,\ldots,\A_k)   \ee
where the $B_k$ are the Bell polynomials, which are defined as
\be e^{\sum_{n=1}^{\infty}A_{n}\frac{t^n}{n!}}=\sum_{k=0}^{\infty}\frac{B_k t^k}{k!}\ee
the first $\A_n$  that can be obtained from direct decomposition of the transfer matrix for low sizes read:
\be  \A_2=-(2L+1)\uno     \ee
\be  \A_4=-6(6L+1)\uno \ee
\be \A_6=-60(40L-4)\uno   \ee
\be \A_8=-(352800 L - 115920)\uno \ee
\be \A_{10}=-(91445760 L - 48625920)\uno \ee
\be \A_{12}=-(36883123200 L - 26424921600)\uno \ee
\be \A_{14}=-(21371135385600 L - 18955051315200)\uno \ee

\be \A_1=\sqrt{2}\sum_{i=1}^{2L+2}\e_i-(2L+2)\uno \ee
\be  \A_3=\HH_3+6\A_1-2\B_1+4\uno \ee
\be  \A_5=6\HH_5+60\HH_3-6[\A_1,[\A_1,\B_1]]+240\A_1-96\B_1-48\B_2+288\uno   \ee
\be \begin{split} \A_7&=90\HH_7+1260\HH_5+7560\HH_3-90[\A_3,[\A_1,\B_1]]-1080[\A_1,[\A_1,\B_1]]-180[\A_1,[\A_1,\B_2]]+\\&+25200\A_1-9360\B_1-9360\B_2-2160\B_3+41760\uno \end{split}  \ee
\be \begin{split} \A_9&=2520\HH_9+45360\HH_7+362880\HH_5+1693440\HH_3-420[\A_5,[\A_1,\B_1]]-25200[\A_3,[\A_1,\B_1]]+\\&-5040[\A_3,[\A_1,\B_2]]-342720[\A_1,[\A_1,\B_1]]-100800[\A_1,[\A_1,\B_2]]-20160[\A_1,[\A_1,\B_3]]\\&+5080320\A_1-1532160\B_1-2499840\B_2-1048320\B_3-161280\B_4+10483200\uno \end{split}  \ee
\be \begin{split} \A_{11}&=113400\HH_{11}+2494800\HH_9+24948000\HH_7+149688000\HH_5+598752000\HH_3-1260[\A_7,[\A_1,\B_1]]\\&-189000[\A_5,[\A_1,\B_1]]-37800[\A_5,[\A_1,\B_2]]-9979200[\A_3,[\A_1,\B_1]]-3628800[\A_3,[\A_1,\B_2]]\\&-907200[\A_3,[\A_1,\B_3]]-167832000[\A_1,[\A_1,\B_1]]-62596800[\A_1,[\A_1,\B_2]]+\\&-17236800[\A_1,[\A_1,\B_3]]-1814400[\A_1,[\A_1,\B_4]]+1676505600\A_1\\&-355622400\B_1-954374400\B_2-555206400\B_3-156038400\B_4-18144000\B_5+4078771200\uno \end{split}  \ee
\be \begin{split} \A_{13}&=7484400\HH_{13}+194594400 \HH_{11}+2335132800\HH_9+17124307200\HH_7+85621536000\HH_5+\\&+308237529600\HH_3-2970[\A_9,[\A_1,\B_1]]-831600[\A_7,[\A_1,\B_1]]-166320[\A_7,[\A_1,\B_2]]+\\&-109771200[\A_5,[\A_1,\B_1]]-39916800[\A_5,[\A_1,\B_2]]-9979200[\A_5,[\A_1,\B_3]]+\\&-5688144000[\A_3,[\A_1,\B_1]]-2634508800[\A_3,[\A_1,\B_2]]-1017878400[\A_3,[\A_1,\B_3]]+\\&-119750400[\A_3,[\A_1,\B_4]]-114960384000[\A_1,[\A_1,\B_1]]-49456915200[\A_1,[\A_1,\B_2]]+\\&-15926803200[\A_1,[\A_1,\B_3]]-3113510400[\A_1,[\A_1,\B_4]]-359251200[\A_1,[\A_1,\B_5]]+\\&+821966745600\A_1-92926310400\B_1-503909683200\B_2-366915225600\B_3-142742476800\B_4+\\&-30656102400\B_5-2874009600\B_6+2280047616000 \uno \end{split}  \ee
\be \begin{split} \A_{15}&=681080400\HH_{15}+20432412000\HH_{13}+286053768000 \HH_{11}+2479132656000\HH_9+\\&+14874795936000\HH_7+65449102118400\HH_5+218163673728000\HH_3-6006[\A_{11},[\A_1,\B_1]]+\\&-2702700[\A_9,[\A_1,\B_1]]-540540[\A_9,[\A_1,\B_2]]-665945280[\A_7,[\A_1,\B_1]]+\\&-242161920[\A_7,[\A_1,\B_2]]-60540480[\A_7,[\A_1,\B_3]]-84453969600[\A_5,[\A_1,\B_1]]+\\&-39956716800[\A_5,[\A_1,\B_2]]-15437822400[\A_5,[\A_1,\B_3]]-1816214400[\A_5,[\A_1,\B_4]]+\\&-4544168428800[\A_3,[\A_1,\B_1]]-2332019289600[\A_3,[\A_1,\B_2]]-1100625926400[\A_3,[\A_1,\B_3]]+\\&-250637587200[\A_3,[\A_1,\B_4]]-32691859200[\A_3,[\A_1,\B_5]]-103981906828800[\A_1,[\A_1,\B_1]]+\\&-49669831411200[\A_1,[\A_1,\B_2]]-17980522560000[\A_1,[\A_1,\B_3]]-4729422297600[\A_1,[\A_1,\B_4]]+\\&-893577484800[\A_1,[\A_1,\B_5]]-65383718400[\A_1,[\A_1,\B_6]]+560992303872000\A_1+\\&-5317875763200\B_1-354379753728000\B_2-306780406732800\B_3-148115916748800\B_4+\\&-44373750220800\B_5-7758867916800\B_6-610248038400\B_7+ 1734673638297600\uno \end{split}  \ee
where we have used the following notation:
\be \HH_3=\sqrt{2}\sum_{k=1}^{2L}[\e_i,[\e_{i+1},\e_{i+2}]]   \ee
\be \HH_5=\sqrt{2}\sum_{k=1}^{2L-2}[\e_i,[\e_{i+1},[\e_{i+2},[\e_{i+3},\e_{i+4}]]]]   \ee
\be \HH_7=\sqrt{2}\sum_{k=1}^{2L-4}[\e_i,[\e_{i+1},[\e_{i+2},[\e_{i+3},[\e_{i+4},[\e_{i+5},\e_{i+6}]]]]]]   \ee
\be  \B_1=\sqrt{2}(\e_{2}+\e_{2L+2})  \ee
\be  \B_2=\sqrt{2}(\e_{3}+\e_{2L+1})  \ee
\be  \B_3=\sqrt{2}(\e_{4}+\e_{2L})  \ee
We call the ${\bf A}_{2n-1}$ the Lattice Local Integrals of Motion (IOM), since they play a completely analogous role as the ${\bf I}_{2n-1}$ of \cite{blz}, as we shall realize even better in the next section.\\
The ${\bf A}_{2n-1}$ form a maximal Abelian subalgebra (of hermitean operators) of the enveloping algebra of the TL algebra, in the same way as the CFT Local IOM ${\bf I}_{2n-1}$ span the maximal abelian subalgebra (of hermitean operators) of the enveloping algebra of the Virasoro algebra. As remarked in \cite{saleur} the 2 enveloping algebras should become isomorphic in a suitably defined $L\to\infty$ limit.\\

\section{Lattice Local Integrals of Motion}
In this section we want to obtain the eigenvalues of the Lattice Integrals of Motion which we have characterized in the previous section in terms of the Temperley-Lieb algebra.\\ 
Starting from the following expansion:
\be  \log T(x)=+\sum_{m=1}^{\infty}\frac{A_n}{n!(2x)^m} \ee
we readily obtain the following formulas for the eigenvalues of the ${\bf A}_{2n-1}$:
\be A_{2n}=2^{2n}(2n-1)!\sum_{k=1}^{L+1}\sin^{2n}(t_k)   \ee
\be A_{2n-1}=2^{2n-1}(2n-2)!(-2\sum_{k\in\mathcal{P}}\sin^{2n-1}(t_k) + \sum_{k=1}^{L+1}\sin^{2n-1}(t_k)  ) \ee
one then observes that the sum over sines in the previous equation can be evaluated analytically thus yielding:
\be \begin{split}  \sum_{k=1}^{L+1}\sin^{2n-1}(t_k)  &=\frac{1}{2^{2n-1}}\sum_{k=0}^{n-1}(-1)^{n-k-1}\binom{2n-1}{k}\csc\Big((2(n-k)-1)\frac{\pi}{4L+6}\Big)\cdot\\&\cdot \Big(1-\sin\Big(\pi\frac{4(n-k)L+4(n-k)+1}{4L+6}\Big)\Big)\end{split} \ee
One then wants to expand the full eigenvalues $A_{2n-1}$ in powers of $4L+6$, for this reason one introduces the following coefficients:
\be \mathcal{C}_{n,m}=\frac{1}{2^{2n-2}}\sum_{k=0}^{n-1}\binom{2n-1}{k}(-1)^{n-k-1}\Big(2(n-k)-1\Big)^{2m-1} \ee
in terms of which one has the following expansions
\be  \sum_{k=1}^{L+1}\sin^{2n-1}(t_k) =2^{2n-3}\frac{(n-1)!^2}{(2n-1)!}\frac{4L+6}{\pi} -\frac{1}{2}+\sum_{m=n}^{\infty}2^{2m-1}\mathcal{C}_{n,m}\frac{(-1)^m}{(2m)!}\Big(\frac{\pi}{4L+6}\Big)^{2m-1}B_{2m}\Big(\frac{1}{2}\Big)   \ee
\be  -2 \sin^{2n-1}(t_k) =\sum_{m=n}^\infty \mathcal{C}_{n,m}\frac{(-1)^m}{(2m)!}\Big(\frac{\pi}{4L+6}\Big)^{2m-1}4m(2k-1)^{2m-1} \ee
where $B_{n}(z)$ are the Bernoulli polynomials. Now, using this results one can obtain that
\be\begin{split} &(-2\sum_{k\in\mathcal{P}}\sin^{2n-1}(t_k) + \sum_{k=1}^{L+1}\sin^{2n-1}(t_k)  ) = 2^{2n-3}\frac{(n-1)!^2}{(2n-1)!}\frac{4L+6}{\pi} -\frac{1}{2}+\\&+\sum_{m=n}^{\infty}2^{2m-1}\mathcal{C}_{n,m}\frac{(-1)^m}{(2m)!}\Big(\frac{\pi}{4L+6}\Big)^{2m-1}\Big(4m\sum_{k\in \mathcal{P}}(k-\frac{1}{2})^{2m-1}+B_{2m}\big(\frac{1}{2}\big)\Big) \end{split} \ee
in which we recognize that the following piece
\be \Big(4m\sum_{k\in \mathcal{P}}(k-\frac{1}{2})^{2m-1}+B_{2m}\big(\frac{1}{2}\big)\Big) =\alpha_m I_{2m-1}(\mathcal{P})  \ee
is proportional to the Local Integrals of Motion of the underlying CFT \cite{nigro}, where
\be  \alpha_n= -\sqrt{\pi}\frac{n(2n-1)3^n\Gamma(4n-1)}{2^{2n-2}n!\Gamma\big(3n-\frac{1}{2}\big)}                  \ee
\be\begin{split} &A_{2n-1}= A_{2n-1}^{div}+2^{2n-1}(2n-2)!\sum_{m=n}^{\infty}\mathcal{C}_{n,m}\frac{(-1)^m}{(2m)!}\Big(\frac{\pi}{2L+3}\Big)^{2m-1}\alpha_m I_{2m-1}(\mathcal{P}) \end{split} \ee
where the divergent part is defined as:
\be A_{2n-1}^{div}= 2^{4(n-1)}\frac{(n-1)!^2}{(2n-1)}\frac{4L+6}{\pi} -2^{2(n-1)}(2n-2)!\ee
notice that the first relevant term in $A_{2n-1}$ aside from the ones appearing in the divergent part is $I_{2n-1}$, thus motivating that the $A_{2n-1}$ being called Lattice Integrals of Motion.\\
We now remark that since ${\cal C}_{n,m}=0$ for $m<n$ the $m$ summation in the formula for $A_{2n-1}$ can also be taken to start from $m=1$.\\
In order to obtain the explicit expression for the $A_{2n}$ we now consider that:
\be \sum_{k=1}^{L+1}\sin^{2n}(t_k)= -\frac{1}{2^{2n}}\Big(\frac{(2n)!}{n!^2}(L+1)-2^{2n-1}+\binom{2n-1}{n}\Big)  \ee
so that we have
\be A_{2n}= -(2n-1)!\Big(\frac{(2n)!}{n!^2}(L+1)-2^{2n-1}+\binom{2n-1}{n}\Big)    \ee
we now decompose the $A_{2n-1}$ as:
\be A_{2n-1}=A_{2n-1}^{div}+\overline{A}_{2n-1}  \ee
and resum the contributions of the divergent part of $A_{2n-1}$ and of $A_{2n}$ so that
\be  \log T(x)= \Big(\sum_{n=1}^\infty \frac{A_{2n}(L)}{(2x)^{2n}(2n)!}+\sum_{n=1}^\infty \frac{A_{2n-1}^{\textrm div}(L)}{(2x)^{2n-1}(2n-1)!}\Big)+\sum_{n=1}^\infty \frac{\overline{A}_{2n-1}}{(2x)^{2n-1}(2n-1)!}      \ee
where we have that
\be \sum_{n=1}^\infty \frac{A_{2n}(L)}{(2x)^{2n}(2n)!}=\frac{2L+3}{2}\log\Big(\frac{1}{2}\big(1+\sqrt{1-\frac{1}{x^2}}\big)\Big)-\log\big(1-\frac{1}{x^2}\big) \ee
\be  \sum_{n=1}^\infty \frac{A_{2n-1}^{\textrm{div}}(L)}{(2x)^{2n-1}(2n-1)!}=\frac{2L+3}{\pi x}\ _3F_2((\frac{1}{2},1,1);(\frac{3}{2},\frac{3}{2});\frac{1}{x^2})-\frac{1}{4}\log\Big(\frac{1+\frac{1}{x}}{1-\frac{1}{x}}\Big)  \ee
furthermore the sum over the $\overline{A}_{2n-1}$ can be rearranged as:
\be \sum_{n=1}^\infty \frac{\overline{A}_{2n-1} }{(2x)^{2n-1}(2n-1)!} =\sum_{m=1}^{\infty}\frac{(-1)^m}{(2m)!}\Big(\frac{\pi}{2L+3}\Big)^{2m-1}\alpha_m I_{2m-1}(\mathcal{P})P_{2m-1}(\frac{1}{x})   \ee
where
\be  P_{2m-1}(z)=\sum_{n=1}^m\frac{z^{2n-1}}{2n-1}\mathcal{C}_{n,m}   \ee
so that the expansion of the logarithm of $T$ takes the form (analogous to the result of \cite{nigropol} for Polymers on the strip ):
\be  \log T(x)= -2(2L+3)f_{bulk}(x)-f_{bou}(x)+ \sum_{m=1}^{\infty}\frac{(-1)^m}{(2m)!}\Big(\frac{\pi}{2L+3}\Big)^{2m-1}\alpha_m I_{2m-1}(\mathcal{P})P_{2m-1}(\frac{1}{x})   \ee
where we have introduced the bulk and boundary free energies:
\be f_{bulk}(x)=-\frac{1}{4}\log\Big(\frac{1}{2}(1+\sqrt{1-\frac{1}{x^2}})\Big)-\frac{1}{2\pi x}\ _3F_2((\frac{1}{2},1,1);(\frac{3}{2},\frac{3}{2});\frac{1}{x^2})  \ee
and
\be  f_{bou}(x)=\log\Big(1-\frac{1}{x^2}\Big)+\frac{1}{4}\log\Big(\frac{1+\frac{1}{x}}{1-\frac{1}{x}}\Big)  \ee
and by taking into account the relation:
\be \int_0^{\frac{\pi}{2}}\log(\frac{1}{\sin(t)}+\frac{1}{x})dt=\frac{1}{x} \ _3F_2((\frac{1}{2},1,1);(\frac{3}{2},\frac{3}{2});\frac{1}{x^2})+\frac{\pi}{2}\log(1+\sqrt{1-\frac{1}{x^2}})  \ee
one gets
\be f_{bulk}(x)=-\frac{1}{2}\Big(\int_0^{\frac{\pi}{2}}\log(\frac{1}{\sin(t)}+\frac{1}{x})dt+\log\sqrt{2}\Big)  \ee
which is the well known formula for the bulk free energy of the Ising model.\\
Notice also that the same kind of expansion has already been obtained for Critical Dense Polymers  on the strip in \cite{nigropol}.

\section{Characters and Operator Content}
Let $p$ be the length of the partition ${\cal P}$, one then defines the finitized characters to be
\be   \chi_+^L(q)=\sum_{\mathcal{P}, \ \textrm{odd}\ p}q^{\sum_{j\in\mathcal{P}}\big(j-\frac{\delta}{2}\big)}   \ee
\be   \chi_{-}^L(q)=\sum_{\mathcal{P}, \ \textrm{even}\ p}q^{\sum_{j\in\mathcal{P}}\big(j-\frac{\delta}{2}\big)}   \ee
which, by defining the q-Binomial
\be  \binom{n}{m}_q=\prod_{i=0}^{m-1}\frac{(1-q^{n-i})}{(1-q^{i+1})}   \ee
can also be expressed in the following Fermionic form:
\be   \chi_{+}^L(q)=\sum_{m=0}^{\lfloor\frac{L+1}{2}\rfloor}q^{2m^2}\binom{L+1}{2m}_q       \ee
\be   \chi_{-}^L(q)=\sum_{m=1}^{\lfloor\frac{L+1}{2}\rfloor+1}q^{2m^2-2m+\frac{1}{2}}\binom{L+1}{2m-1}_q       \ee
alternatively we also have the following Bosonic forms for the finitized characters:
\be  \chi_{+}^L(q)=\frac{1}{2}\Big(\prod_{k=1}^{L+1}(1+q^{k-\frac{1}{2}})+\prod_{k=1}^{L+1}(1-q^{k-\frac{1}{2}})\Big)   \ee
\be  \chi_{-}^L(q)=\frac{1}{2}\Big(\prod_{k=1}^{L+1}(1+q^{k-\frac{1}{2}})-\prod_{k=1}^{L+1}(1-q^{k-\frac{1}{2}})\Big)   \ee
We remark that these finitized characters have been already object of extensive study, and are found, in a slightly less symmetric form for example in \cite{ret1}.\\
The characters with the $+$ subscript enumerate the states corresponding to the $b=+1$ boudary condition whereas the subscript $-$ corresponds to $b=-1$.
By use the above formulas one can see that the partition functions behave as:
\be  Z=\textrm{Tr}\Big(\T^M\Big)\sim Z_{\textrm{div}}(L,M)q^{-\frac{1}{48}}\chi_{\pm}^L(q)   \ee
where
\be   q=e^{-\pi\frac{M}{(2L+3)x}}  \ee
and:
\be  \log Z_{\textrm div}= -2M(2L+3)f_{bulk}(x)-Mf_{bou}(x)  \ee
We now take a look at the expansions of the $L=\infty$ characters:
\be \chi_{+}^\infty(q)=1+q^2+q^3+2q^4+2q^5+3q^6+3q^7+5q^8+5q^9+7q^{10}+8q^{11}+11q^{12}+12q^{13}+\ldots   \ee
\be \chi_{-}^\infty(q)=q^{\frac{1}{2}}+q^{\frac{3}{2}}+q^{\frac{5}{2}}+q^{\frac{7}{2}}+2q^{\frac{9}{2}}+2q^{\frac{11}{2}}+3q^{\frac{13}{2}}+4q^{\frac{15}{2}}+5q^{\frac{17}{2}}+6q^{\frac{19}{2}}+8q^{\frac{21}{2}}+9q^{\frac{23}{2}}+\ldots    \ee
we notice that:
\be  \chi_{+}^\infty(q)=\chi_{0}(q)    \ee
\be  \chi_{-}^\infty(q)=\chi_{\frac{1}{2}}(q)    \ee
whre $\chi_0$ and $\chi_{\frac{1}{2}}$ are the respectively the characters of the irreducible $h=0$, $h=1/2$ representations of the Virasoro algebra.\\
The states belonging to  the $(0)_{+}$ and $(\frac{1}{2})_{-}$ representations, and labelled by a partition $\mathcal{P}$, will all be of the form:
\be  \big|\mathcal{P}\big>=\psi_{-\frac{2k_p-1}{2}}\ldots\psi_{-\frac{2k_1-1}{2}}\big|0\big>  \ee
with $p$ respectively even or odd.\\
The fermion field is expanded in this case as:
\be  \psi(z)=\sum_{m\in\mathbb{Z}+\frac{1}{2}}\frac{\psi_{m}}{z^{m+\frac{1}{2}}}  \ee
where the fermi modes satisfy in all cases the following anticommutation relations:
\be  \{\psi_n,\psi_m\}=\delta_{n+m,0}    \ee
All the above conformal characters and fermionic decompositions are well known and appear in many works, we point out here for example \cite{paths}\cite{ginsparg} as particularly understandable.

\section{Acknowledgements}
The author acknowledges financial support from Fondo Sociale Europeo (Regione Lombardia), through the grant ÒDote ricercaÓ.\\

\end{document}